\documentclass{aa}

\usepackage{graphicx}
\usepackage{tabularx}
\usepackage{array}
\usepackage{txfonts}
\usepackage{xcolor}
\usepackage{lipsum}
\usepackage{siunitx}
\usepackage{subcaption}
\usepackage{lscape}
\usepackage{placeins}
\usepackage[colorlinks=true,allcolors=blue]{hyperref}

\graphicspath{{images}}

\DeclareMathAlphabet{\pazocal}{OMS}{zplm}{m}{n}

\begin{document}

\title{Giant exoplanets are not fully mixed}

\author{Chantal Hemmann\inst{1,2}\fnmsep\thanks{Corresponding author: chantalhemmann@gmail.com}
    \and Simon M\"{u}ller\inst{2}
    \and Ravit Helled\inst{2}
    }

\institute{
    Dept. of Physics and Astronomy, University of British Columbia, Canada
    \and Dept. of Astrophysics, University of Zurich, Winterthurerstr. 190, 8057 Zurich, Switzerland
}

\date{Received Month Day, Year}

\abstract
% context heading
{The interior structure and bulk composition of giant planets are not
directly observable and must be inferred from models. Under the common assumption of a well-mixed,
adiabatic envelope, the measured atmospheric metallicity is taken as a
proxy for the metallicity of the entire planetary envelope, and hence as a handle
on the planet's heavy-element budget. JWST now provides precise
atmospheric metallicities for a growing number of warm giants, making it
possible to test this assumption for exoplanets for the first time.}
% aims heading
{We quantify the difference between the envelope and bulk metallicities 
of warm giant exoplanets to assess the evidence for 
compositional stratification as opposed to homogeneity in their interiors.}
% methods heading
{We assembled a sample of eleven warm giant exoplanets
with atmospheric metallicities derived from published JWST retrievals. 
We computed tailored grids of interior and
thermal evolution models for each planet and performed MCMC retrievals to
infer the bulk metallicity consistent with the measured mass, radius,
system age, and atmospheric metallicity.}
% results heading
{We find that the envelope metallicities are smaller than the bulk metallicities 
throughout the sample, with mixing ratios ranging from about $0.02$ to $0.90$.
Eight of the eleven planets have mixing ratios smaller than $0.50$, and ten
of eleven are inconsistent with a fully mixed interior to within one
sigma. We tentatively identify a statistically significant anti-correlation between
planetary mass and envelope metallicity, but find no correlation between
envelope and bulk metallicity, nor between envelope and host-star
metallicity.}
% conclusions heading
{Atmospheric metallicity is not a reliable proxy for the bulk composition in
warm giant planets, and incomplete mixing (possibly, composition gradients) appears to be common among the
planets currently accessible to JWST characterization. Bulk composition
estimates that assume a homogeneous envelope substantially
underestimate the total heavy-element mass. That the solar-system giants
are unremarkable within this sample suggests dilute or partially mixed
interiors may be a generic outcome of giant planet formation.}


\maketitle

\nolinenumbers
\section{Introduction}

Understanding the processes governing the formation and evolution of giant planets, and resolving the associated uncertainties, has long been a fundamental goal of planetary science. Key properties governing the formation and evolution of planets, including their interior structure and bulk composition, are not directly accessible to observation and must therefore be inferred from theoretical models. These simulations are highly degenerate, and inferred quantities are greatly sensitive to the particular selection of model assumptions, including the equation of state \citep[EOS; e.g.][]{baraffe_structure_2008,vazan_effect_2013,muller_theoretical_2020,howard_accounting_2023}, atmospheric model and metallicity \citep{poser_effect_2019,muller_synthetic_2021,poser_effect_2024}, and interior architecture---for example whether the heavy element content is concentrated in a compact core, or distributed in a broader dilute ("fuzzy") structure \cite[][]{howard_accounting_2023, peerani_how_2026}.

The unprecedented accuracy in atmospheric characterization courtesy of the \textit{James Webb} Space Telescope \citep[JWST;][]{gardner_james_2006, barstow_transit_2015} supplies spectroscopic constraints for inferring the atmospheric composition. Probing the upper layers of exoplanetary atmospheres ($\sim$ 1-100 mbar), transmission and emission spectra have been measured with JWST's Near Infrared Spectrograph (NIRSpec; $0.6-5.8$ µm), Near Infrared Camera (NIRCam; $2.4-5$ µm), and Mid-Infrared Instrument (MIRI; $5-28$ µm). These spectra are analyzed via atmospheric retrieval codes (e.g., POSEIDON \citet{macdonald_poseidon_2023}, PICASO \citet{mukherjee_picaso_2023}) to derive an atmospheric metallicity $\text{log}_{10}$M/H = [M/H] from molecular absorption features. \footnote{In this work, we use the standard astrophysical definition of metallicity, where M represents all elements heavier than helium.} In some cases the carbon-to-oxygen (C/O) ratio is also measured, potentially providing a complementary probe of formation location and accretion history \citep{mordasini_imprint_2016,madhusudhan_atmospheric_2017,turrini_tracing_2021,schneider_how_2021,pacetti_chemical_2022}.

Under the assumption of a well-mixed, adiabatic envelope, the measured atmospheric metallicity can be taken as a proxy for the metallicity in the entire envelope ($Z_\text{env}$) for comparison against the planet's total bulk metallicity ($Z_\text{bulk}$), which may contain contributions from a central core.
Such a comparison is key to distinguishing formation pathways: the core accretion model generally predicts a metal enrichment relative to the host star \citep[e.g.][]{thorngren_massmetallicity_2016,ikoma_formation_2025,chachan_revising_2025}, whereas the disk instability model predicts near-stellar abundances unless there is subsequent external enrichment \citep{helled_heavy-element_2009,boley_possibility_2010,humphries_changes_2018}.
A significant discrepancy between $Z_\text{env}$ and $Z_\text{bulk}$ in this case would indicate a inhomogeneous planet, where the heavy element content is concentrated in the deeper interior, whether as a primordial compositional gradient or the remnant of a compact core.
Alternatively, a similar $Z_\text{env}$ and $Z_\text{bulk}$ would imply a planet is well-mixed, with a largely homogeneous composition.
As planets and their host stars form from a common proto-stellar cloud, comparisons between planetary and stellar metallicities further constrain accretion efficiency and metal retention \citep{hasegawa_planet_2014,teske_metal-rich_2019,thorngren_connecting_2019,swain_planet_2024}.

With the exception of the Solar System gas giants, comparisons between atmospheric metallicities and bulk metallicities in warm giant exoplanets have historically been hampered by the lack of observational constraints on atmospheric composition. With the advent of JWST and, in the near future, ARIEL \citep{tinetti_chemical_2018}, increasingly precise constraints on atmospheric metallicities are becoming possible. This enables, for the first time, systematic comparisons between atmospheric and bulk metallicities in warm giant exoplanets, providing a powerful means of assessing the degree of compositional mixing between their atmospheres and interiors. 

In this work, we created the first dataset that compares bulk and atmospheric metallicites of giant planets: We used the giant-planet interior and evolution models to constrain the bulk metallicity of eleven warm giant exoplanets with JWST-derived atmospheric metallicities from published spectral retrievals. We benchmark our findings against Jupiter and Saturn, whose bulk metallicities are independently constrained by Juno and Cassini gravity field measurements \citep[e.g.,][]{miguel_interior_2023, helled_giant_2024}. Our paper is organized as follows: The evolution model, planet sample, and interior-retrieval method are described in Section \ref{sec:methods}. The main results are presented in Section \ref{sec:results}, where we also discuss a tentative statistical analysis. We discuss our results in Section \ref{sec:discussion}, and present our conclusions in Section \ref{sec:conclusions}. The host-star parameters of our planet sample are tabulated in Appendix \ref{sec:appendix_1}.

\section{Methods}\label{sec:methods}
In this section, we first describe evolution models used in this work. Then, we present the sample of planets consisting of warm giant exoplanets with measured atmospheric metallicities, and describe the method used to infer the bulk metallicity of these planets.

\subsection{Interior and evolution models}

\begin{table*}[ht!]
\renewcommand{\arraystretch}{1.3}
\caption {Physical characteristics of planets in the sample, in order of increasing mass.}
\label{tab:planet_sample} 
\centering
\begin{tabularx}{\textwidth}{l*{6}{>{\centering\arraybackslash}X}}
\hline            
Planet & Mass ($M_{\text{Jup}}$) & Radius ($R_{\text{Jup}}$) & Age (Gyr) & $T_{\text{eq}}$ (K) & [M/H] & C/O \\
\hline
HAT-P-26 b & 0.06$\pm$0.01 & 0.65$\pm$0.06 & 5.9$^{+4.8}_{-3.8}$ & 1000$\pm 25$ & 1.04$^{+0.07}_{-0.14}$ & 0.14$^{+0.21}_{-0.08}$ \\
WASP-107 b & 0.10$\pm$0.01 & 0.94$\pm$0.03 &  3.4$^{+0.7}_{-0.7}$ & 673$\pm$147 & 1.09$^{+0.17}_{-0.07}$ & 0.33$^{+0.06}_{-0.05}$ \\
TOI-3984 A b & $0.14\pm0.03$ & $0.17\pm0.02$ & 0.7-5.1 & $563\pm15$ & $-0.12^{+0.68}_{-0.86}$ & -- \\
TOI-199 b & 0.17$\pm$0.02 & 0.81$\pm$0.01 & 0.8$^{+1.2}_{-0.6}$ & 350$\pm$25 & 1.11$^{+0.85}_{-1.15}$ & -- \\
HAT-P-18 b  & 0.20$\pm$0.02 & 1.00$\pm$0.05 & $12.4^{+1.4}_{-6.4}$ & 841$\pm$15 & -0.77$^{+0.47}_{-0.40}$ & 0.21$^{+0.23}_{-0.10}$ \\
HAT-P-12 b & 0.21$\pm$0.01 & 0.95$\pm$0.06 & 2.5$^{+2.0}_{-2.0}$ & 975$\pm$8 & 0.88$^{+0.64}_{-0.91}$ & 0.58$^{+0.20}_{-0.23}$ \\
WASP-69 b  & 0.26$\pm$0.02 & 1.00$\pm$0.02 & 1.8$^{+0.8}_{-0.8}$ & 971$\pm$33 & 0.96$^{+0.20}_{-0.17}$ & 0.75$^{+0.19}_{-0.10}$ \\
% HATS-6 b & 0.36$\pm$0.09 & 1.01$\pm$0.02 & 8.1$^{+4.3}_{-4.3}$ & 740$\pm$12 & $-2.00^{+0.20}_{-0.20}$ & -- \\
HATS-75 b & 0.49$\pm$0.04 & 0.88$\pm$0.01 & 14.9$^{+3.3}_{-4.3}$ & 772$\pm$2 & $-1.74^{+0.92}_{-0.76}$ & -- \\
WASP-80 b  & 0.54$\pm$0.04 & 0.95$\pm$0.03 & 7.0$^{+7.0}_{-7.0}$ & 825$\pm$19 & 0.55$^{+0.12}_{-0.10}$ & 0.48$^{+0.06}_{-0.07}$ \\
% TOI-3714 & $0.70\pm0.03$ & $1.01\pm0.03$ & \textbf{0.7-5.1}  & $750\pm20$& $-1.30^{+0.30}_{-0.30}$& -- \\
TOI-5205 b & 1.08$\pm$0.06 & 1.04$\pm$0.03 & 6.0$^{+4.0}_{-4.0}$ & 737$\pm$15 & -1.90$^{+0.20}_{-0.20}$ & --\\
%TOI-5205 b & 1.08$\pm$0.06 & 1.04$\pm$0.03 & 6.0$^{+4.0}_{-4.0}$ & 737$\pm$15 & -1.92$^{+0.11}_{-0.05}$ & 1.30$^{+0.40}_{-0.40}$ \\
HD-80606 b  & 4.16$\pm$0.01 & 1.03$\pm$0.02 & 5.9$^{+1.6}_{-2.0}$ & 405$\pm$7 & -0.17$^{+0.31}_{-0.31}$ & 0.49$^{+0.15}_{-0.15}$ \\
%TOI-5293 A b & 0.50$\pm$0.07 & 0.98$\pm$0.04 & 3.00$^{+2.00}_{-2.00}$ & 676$\pm$20 & $-1.03^{+0.53}_{-0.44}$ & -- \\
\hline

\end{tabularx}
\caption*{$T_{\text{eq}}$ is the zero Bond albedo equilibrium temperature. References: \textbf{HAT-P-26 b}: \cite{mancini_gaps_2022}; \cite{gressier_jwst-tst_2025} (Teq uncertainties estimated) - \textbf{WASP-107 b}: \cite{piaulet_wasp-107bs_2021, howard_planet_2025}; \cite{welbanks_high_2024} - \textbf{TOI-199 b}: \cite{bello-arufe_methane_2025} \cite{hobson_toi-199_2023} (Teq uncertainties estimated) - \textbf{HAT-P-18 b}: \cite{bonomo_gaps_2017}; \cite{esposito_gaps_2014}; \cite{changeat_cloud_2025} (Free retrieval) - \textbf{HAT-P-12 b}: \cite{ozturk_new_2019}; \cite{ment_radial_2018}; \cite{crouzet_detection_2025} - \textbf{WASP-69 b}: \cite{allart_nirps_2025}; \cite{schlawin_multiple_2024} (Cloud layer retrieval) - \textbf{WASP-80 b}: \cite{bonomo_gaps_2017}; \cite{wiser_precise_2025} - \textbf{TOI-5205 b}: \cite{canas_gems_2025} (PLATON, chemical equilibrium); \cite{kanodia_toi-5205_2023} - \textbf{HD 80606 b}: \cite{sikora_seasonal_2025}; \cite{bonomo_gaps_2017}; \cite{pearson_utilizing_2022}; \cite{southworth_homogeneous_2011} - \textbf{HATS-75 b}: \cite{jordan_hats-74ab_2022}; \cite{ashtari_gems_2026} - \textbf{TOI-3984 A b}: \cite{canas_toi-3984_2023}; Wallack et al. (in prep.)}
\end{table*}

% \textbf{HATS-6 b}: \cite{hartman_hats-6b_2015}; Guzman Caloca et al. (in rev.)
% \textbf{TOI-3714 b}: \cite{canas_toi-3714_2022}; Delamer et al. (in prep.)
% \textbf{TOI-5293 A b}: \cite{weisserman_aligned_2025}; Kanodia et al (2026)

We used the \textsc{GASTLI} code \citep[][]{acuna_gastli_2024} to model the interiors and evolution of giant exoplanets. The code self-consistently solves the coupled standard structure equations (hydrostatic equilibrium and mass conservation) and cooling equations governing entropy evolution to model a planet's structural evolution over time. The model assumes a three-layer planetary structure consisting of a core (50\% water and 50\% silicates by weight), envelope (hydrogen, helium, and water), and atmosphere of variable metal content, and that there is no mixing between the layers. The atmosphere is based on a grid of self-consistent one-dimensional chemical equilibrium atmospheric models. The SESAME \cite{sesame_eos, miguel_jupiters_2022} dry sand and \cite{chabrier_new_2021} EOS tables are used for the density and thermodynamic properties of silicates and hydrogen-helium respectively. The interface between the atmosphere and the fully-adiabatic envelope was set at a pressure of $P = 1000$ bar. When calculating the thermal evolution, we assumed a hot start to the planet with an initial entropy of $12 \, k_{\text{B}}/m_{\text{H}}$.
 
Some of these assumptions are clearly simplified, such as the three-layer model: The more modern view of giant planets that includes composition gradients in the deep interior or even the upper envelope \citep[e.g.,][]{wahl_comparing_2017,debras_new_2019,helled_giant_2024,howard_exploring_2023,muller_can_2024}. 
As previous work showed, this and other model assumptions affected the inferred bulk metallicities, which therefore always have an intrinsic theoretical uncertainty \citep{muller_theoretical_2020,howard_giant_2025,peerani_how_2026}. In our following analysis it should therefore be kept in that mind that the true uncertainty on the bulk metallicity is very likely larger than estimated from our statistical method. Despite these uncertainties, however, bulk metallicity inferences provide useful information and can be used to infer trends on a population level or a sample of planets \citep[e.g.,][]{thorngren_massmetallicity_2016,teske_metal-rich_2019,muller_bulk_2025}.

\subsection{Planet Sample}

Of the planetary atmospheric measurements from JWST observations accessible to us, eleven were within the viable ranges of planetary masses and equilibrium temperatures for \textsc{GASTLI}. The physical characteristics of these planets are summarized in Table \ref{tab:planet_sample}. While the majority of these planets orbit G- and K-type stars, three targets (TOI-5205 b, TOI-3984 A b, and HATS-75 b) orbit M-dwarfs. The planets spanned a range of $0.06 M_{\text{Jup}} - 4.16 M_{\text{Jup}}$; most of them are in the sub-Saturn ($M_p \lesssim 0.3 M_{\text{Jup}}$) regime. Hot Jupiters with equilibrium temperatures above 1000 K were excluded to avoid the complexities caused by their unknown inflationary mechanisms \citep[e.g.][]{fortney_hot_2021} and since they were out-of-bounds of the atmospheric grid. Stellar values and additional parameters used for analysis are listed in Appendix \ref{sec:appendix_1}.

\subsection{Statistical Inference}

For each of the eleven planets studied in this work, model parameters were inferred by requiring the thermally evolved model radius at the observed system age to approximately reproduce the observed mass, radius, and atmospheric metallicity, where the atmospheric metallicity corresponds to the retrieved $\log[\text{M/H}]$ from JWST spectroscopy. 

For the statistical inference of model parameters, for each planet we first constructed a tailored grid of interior models. These grids spanned roughly $\pm 3 \sigma$ the reported planetary mass $M_{\text{p}}$ and approximately $\pm 1 \sigma$ the reported atmospheric metallicity $\log[\text{M/H}]$, while the core-mass fraction (CMF) spanned a range from zero up to a customized maximum. We used a fixed zero-albedo equilibrium temperature $T_{\text{eq}}$ and did not include its uncertainties; a simplification that is not expected to significantly impact the results for the deep interior as it is a comparatively well-constrained quantity. Similarly, we fixed the C/O ratio to that derived by the retrieval study. In cases where the reported C/O ratio exceeded the range of atmospheric models or was not provided (e.g. for TOI-199 b), an envelope of solar composition (C/O = 0.55) was assumed based on \cite{asplund_chemical_2009,caffau_solar_2011} solar abundances to avoid extrapolation.

The choice of the atmospheric metallicity is more nuanced: depending on the assumed atmospheric model, retrieval results for the same planet can differ by a factor of several, for example in WASP80-b's case, where cloud-layer and free retrievals yield different [M/H] \citep[see Table 3 in][]{schlawin_multiple_2024}. In cases where multiple retrievals were presented \citep[e.g.][presenting four model setups with four corresponding metallicities and C/O ratios]{schlawin_multiple_2024}, the most plausible scenario as indicated by the respective authors was used for this analysis. 
% The grid range and resulting corner plot for each planet is in Appendix 1.

For each planet, at every grid point thermal evolution sequences were generated over a series of discrete intrinsic temperatures $T_{int}$. These sequences were subsequently re-parameterized in terms of planetary age by interpolating onto a common grid of 256 evenly-spaced points over 0.1 - 14 Gyr. This re-parameterization was motivated by the fact that the intrinsic temperature of a planet is not a directly observable quantity. A more readily available constraint is the stellar age. We therefore used the host star's estimated age as the planet's age, justified by the negligible delay between stellar and planetary formation compared to Gyr evolutionary timescales.

The resulting model outputs were stored as 5-dimensional HDF5 datasets with axes (CMF, $\log[\text{M/H}]$, $M_\text{p}$, $T_{\text{eq}}$, age).
For each grid point, we recorded the derived quantities of interest: planetary radius $R_{\text{p}}$, mass $M_{\text{p}}$, intrinsic temperature $T_{\text{int}}$, envelope metallicity $Z_{\text{env}}$, and bulk metallicity $Z_{\text{bulk}}$. The envelope and bulk metallicites are the mass fractions of heavy elements in the envelope and entire planet, respectively. The bulk metallicity was computed as the as the mass-weighted sum of the heavy-element contributions from the core and the envelope as $Z_{bulk} =\text{CMF} + (1-\text{CMF}) \, Z_{env}$ where core is assumed to consist entirely of heavy elements. The envelope metallicity is an output of the atmospheric models in \textsc{GASTLI}, and is calculated from the input atmospheric log[M/H] using data from \textsc{easyCHEM} \citep{lei_easychem_2025}.

This setup allowed us to construct a numerically efficient forward model by linearly interpolating on the grid of models using the \textsc{scipy.interpolate.RegularGridInterpolator} class. We used these forward models in a Markov Chain Monte Carlo (MCMC) framework with the ensemble sampler \textsc{emcee} \citep{foreman-mackey_emcee_2013} to constrain the planetary parameters. For the ensemble sampler we used 32 walkers and 100,000 steps, and verified convergence by requiring the chain length to exceed 50 times the integrated autocorrelation time for all parameters. 

To constrain the independent parameters to within the bounds of the grid, we used uninformed uniform priors for the core-mass fraction, planetary atmospheric metallicity and mass. For the age, we used an informed uniform prior with lower and upper limits based on stellar age estimates. If the drawn age was older than 13.8 Gyr (the age of the Milky Way), we set it to this value. To constrain the model parameters by the observations, we adopted total likelihood as the product of independent (zero-covariance) Gaussian likelihoods in the observed planetary mass, radius, and atmospheric metallicity, such that the log-likelihood was:

\begin{equation}
    \label{eq:likelihood}
    \ln \mathcal{L} = 
    -\frac{1}{2} \sum_i 
    \left( \frac{d_i}{\sigma_i} \right)^2,
\end{equation}

\noindent
where $d_i$ and $\sigma_i$ are the mean values and standard deviations of the observational data.

\section{Results}\label{sec:results}

We inferred the bulk metallicites as constrained by the planetary masses, radii, ages, and atmospheric metallicites of our planet sample (see Table \ref{tab:planet_sample}) as described in the previous section. The results are summarized in Table \ref{tab:results}, where we list the inferred envelope and bulk metallicities ($Z_\text{env}$ and $Z_\text{bulk}$) as well as their mixing ratios ($Z_\text{env}/Z_\text{bulk}$). For comparison, we also list these values for Jupiter and Saturn.

\begin{table}[ht!]
\renewcommand{\arraystretch}{1.3}
\renewcommand{\tabcolsep}{3pt}
\caption{Interior retrieval results}                 % title of Table
\label{tab:results}    % is used to refer this table in the text
\centering                        % used for centering table
\begin{tabular}{lcccr}      % centered columns (4 columns)
\hline\hline               % inserts double horizontal lines
Planet & $Z_{\text{env}}$ & $Z_{\text{bulk}}$ & $Z_{\text{env}}/Z_{\text{bulk}}$ \\         % table heading
\hline 
% inserts single horizontal line
  HAT-P-26 b & 0.16$^{+0.06}_{-0.06}$ & 0.53$^{+0.04}_{-0.09}$ & 0.32$^{+0.13}_{-0.11}$\\
  WASP-107 b & 0.09$^{+0.04}_{-0.01}$ & 0.14$^{+0.04}_{-0.03}$ & 0.73$^{+0.18}_{-0.18}$\\
  TOI-3984 A b & $0.04^{+0.03}_{-0.02}$ & $0.41^{+0.05}_{-0.05}$ & $0.10^{+0.06}_{-0.06}$ \\
  TOI-199 b & 0.10$^{+0.10}_{-0.05}$ & 0.32$^{+0.15}_{-0.05}$ & 0.30$^{+0.35}_{-0.16}$ \\
  HAT-P-18 b & 0.02$^{+0.01}_{-0.01}$ & 0.10$^{+0.06}_{-0.05}$ & 0.20$^{+0.22}_{-0.09}$\\
  HAT-P-12 b & 0.06$^{+0.04}_{-0.03}$ & 0.14$^{+0.09}_{-0.05}$ & 0.42$^{+0.35}_{-0.24}$ \\
  WASP-69 b & 0.08$^{+0.02}_{-0.02}$ & 0.09$^{+0.02}_{-0.02}$ & 0.89$^{+0.08}_{-0.14}$\\
  % HATS-6 b & $2.5\text{e}\!-\!4$ $^{+39.6\text{e}-4}_{-2.4\text{e}-4}$ & 0.14$^{+0.09}_{-0.08}$ & $0.017$ $^{+0.018}_{0.017}$ \\
  HATS-75 b & $0.003^{+0.002}_{-0.002}$ & 0.20$^{+0.04}_{-0.04}$ & $0.015$ $^{+0.011}_{-0.010}$  \\
  WASP-80 b & 0.07$^{+0.01}_{-0.01}$& 0.13$^{+0.07}_{-0.04}$ & 0.55$^{+0.25}_{-0.21}$ \\
  % TOI-3714 b & $0.005^{+0.002}_
  % {-0.002}$ & $0.15^{+0.08}_{-0.09}$ & $0.031^{+0.037}_{-0.016}$ \\
  TOI-5205 b & $7.5\text{e}\!-\!4 ^{+13.4\text{e}{-4}}_{-7.4\text{e}{-4}}$ & $0.15^{+0.08}_{-0.07}$ & 0.05$^{+0.09}_{-0.07}$\\
  HD-80606 b & 0.02$^{+0.01}_{-0.01}$ & 0.16$^{+0.08}_{-0.07}$ & 0.11$^{+0.13}_{-0.05}$ \\
%  TOI-5293 A b & $1.48\text{e}\!-\!3$ $^{+4.70\text{e}-3}_{-5.06\text{e}-4}$ & 0.02$^{+0.03}_{-0.01}$ & 0.08$^{+0.22}_{-0.06}$ \\
\hline
  Jupiter & 0.02$^{+3.70\text{e}-3}_{-3.30\text{e}-3}$ & 0.06$^{+0.06}_{-0.03}$ & 0.28$^{+0.36}_{-0.14}$  \\
  Saturn & 0.08$^{+0.03}_{-0.02}$ & 0.19$^{+0.02}_{-0.02}$ & 0.42$^{+0.19}_{-0.13}$ \\
\hline
\vspace{0.1 mm}
\end{tabular}
\caption*{The mixing ratio $Z_{\text{env}}\, / \, Z_{\text{bulk}}$ is taken to be the median of the posterior distributions. Estimates for the bulk and atmospheric metallicities for Jupiter and Saturn are from \cite{helled_giant_2024}.}
\end{table}

 \begin{figure*}[t]
    \centering
    \includegraphics[width=0.8\textwidth]{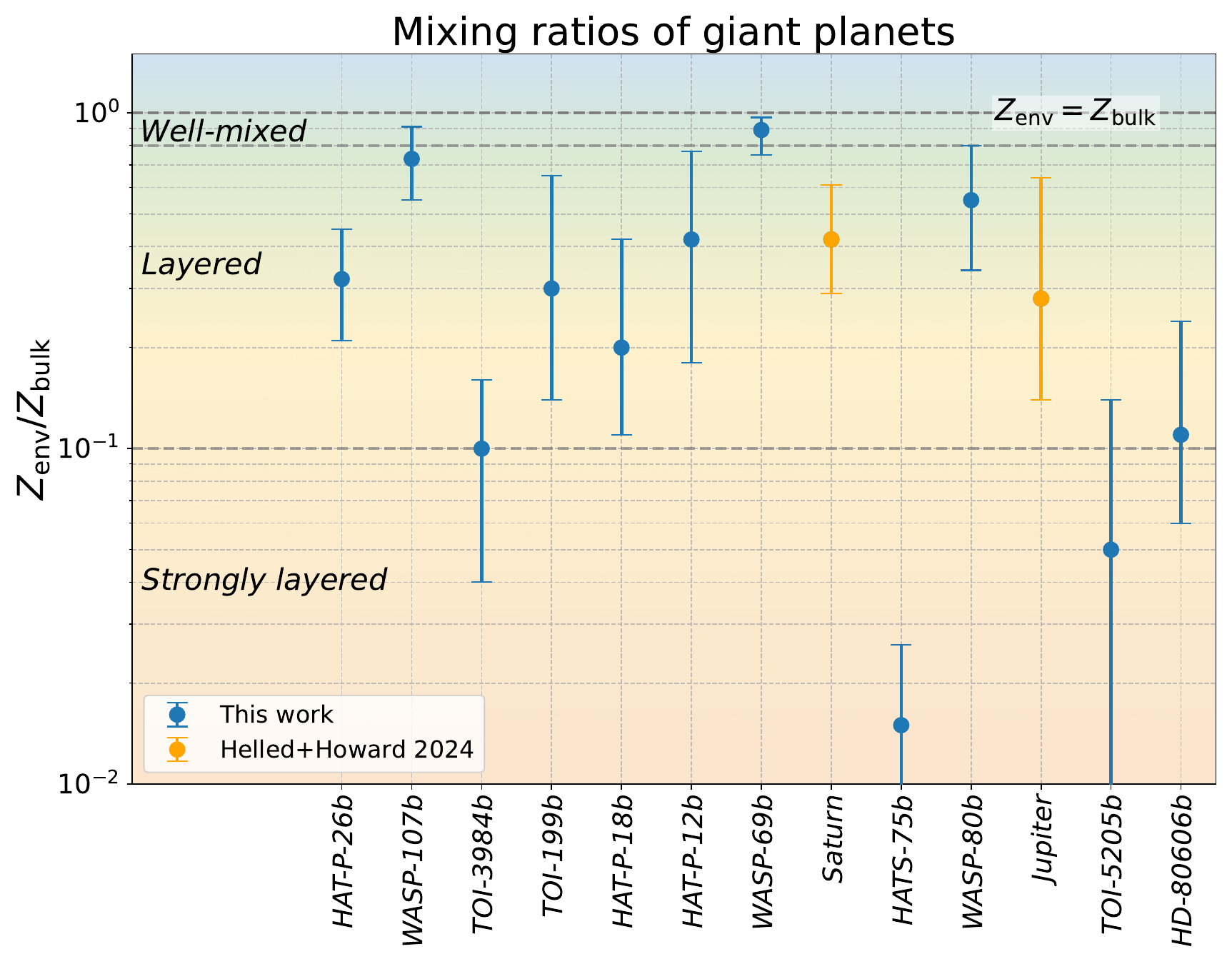}
      \caption{Mixing ratios $Z_\text{env} \, / \, Z_\text{bulk}$ for the eleven exoplanets (blue) in order of ascending mass. Jupiter and Saturn are included for reference based on \cite{helled_giant_2024}. The region below $10^{-2}$ is excluded for readability; for such low ratios, a well-mixed planet can clearly be ruled out. Dashed lines represent approximate boundaries between structural conclusions. A $Z_\text{env} \, / \, Z_\text{bulk}$ > 1 would represent a significantly enriched atmosphere compared to the bulk of the planet.}
         \label{fig:mixing_ratios}
\end{figure*}

Figure \ref{fig:mixing_ratios} shows the mixing ratios of the  eleven planets (plus Jupiter and Saturn). As expected from previous results, we find that there is a large range of inferred bulk metallicites within our sample, namely from $Z_\text{bulk} \approx 0.1 - 0.5$. The envelope metallicites also show a significant scatter from $Z_\text{env} \approx 10^{-3} - 0.16$, and they tend to be significantly lower then the bulk metallicites. This stark contrast between the two metallicites is easily seen in the mixing ratios. The mixing ratios generally have quite large uncertainties, due to both both uncertainties in the measured $\log[\text{M/H}]$ and the inferred $Z_\text{bulk}$. The uncertainties in the latter are primarily driven by uncertainties in the planetary radius, although for some planets, the stellar age is also poorly constrained and can therefore represent an important additional source of uncertainty. 

The mixing ratio can be interpreted as a measure of how well a planet is mixed: A value of one corresponds to a fully mixed planet. Small values suggest that there must be a strong compositional difference between the atmosphere and the planetary bulk. Intermediate values indicate a moderately stratified interior, in which the deeper interior is more metal-rich than the overlying envelope. Values above unity would instead indicate an inverted heavy-element gradient \citep{howard_exploring_2023,muller_can_2024}; however, no planet in our sample exhibits such a configuration. 

We find that the majority of planets lie within the loosely-defined "moderately stratified" zone between mixing ratios of $0.1$ to $0.8$. Jupiter and Saturn, which are expected to have dilute/fuzzy cores \citep[e.g.,][]{wahl_comparing_2017,debras_new_2019,mankovich_diffuse_2021,miguel_jupiters_2022}, also lie within this range \citep{helled_giant_2024}.  WASP-107 b and WASP-80 b are also among the more well-mixed planets in our sample. The mixing ratio of WASP-69 ($Z_\text{env}/Z_\text{bulk}=0.89^{+0.08}_{-0.14}$) stands out as the highest out of all planets in our sample, consistent with near-homogeneity between the envelope and bulk metal enrichment within about one sigma. The two remaining planets HATS-75b and TOI-5205b had the most metal-poor predicted envelopes, and demonstrated the strongest stratification, with $Z_\text{env}/Z_\text{bulk}<0.1$; notably, they both orbit M-dwarf stars. HATS-75b demonstrated an extremely small mixing ratio ($<10^{-3}$) comparatively, indicative of strong metal depletion in their envelopes relative to their bulk composition. Across the planet sample, we consistently find $Z_\text{env} < Z_\text{bulk}$.

\subsection{Preliminary statistical analysis}

\begin{figure*}[t]
  \centering
  \includegraphics[width=0.33\textwidth]{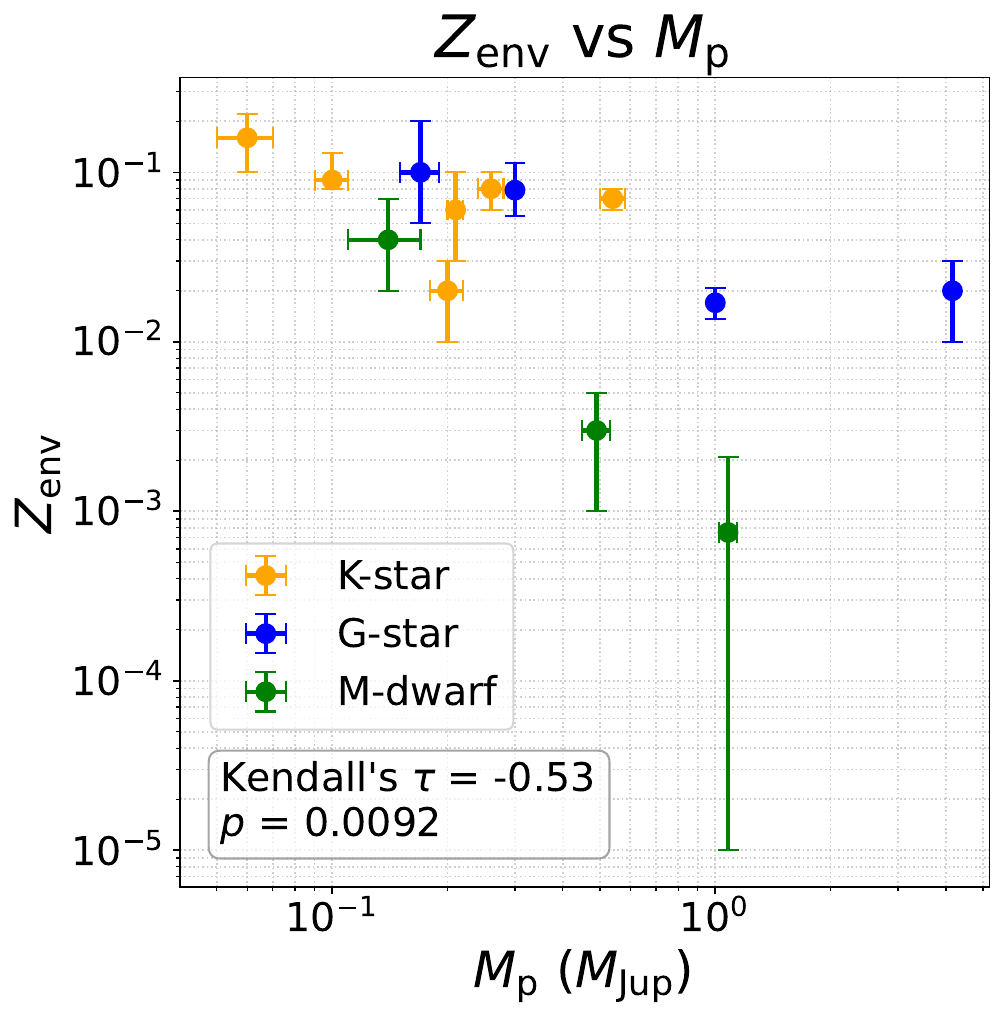}  \includegraphics[width=0.33\textwidth]{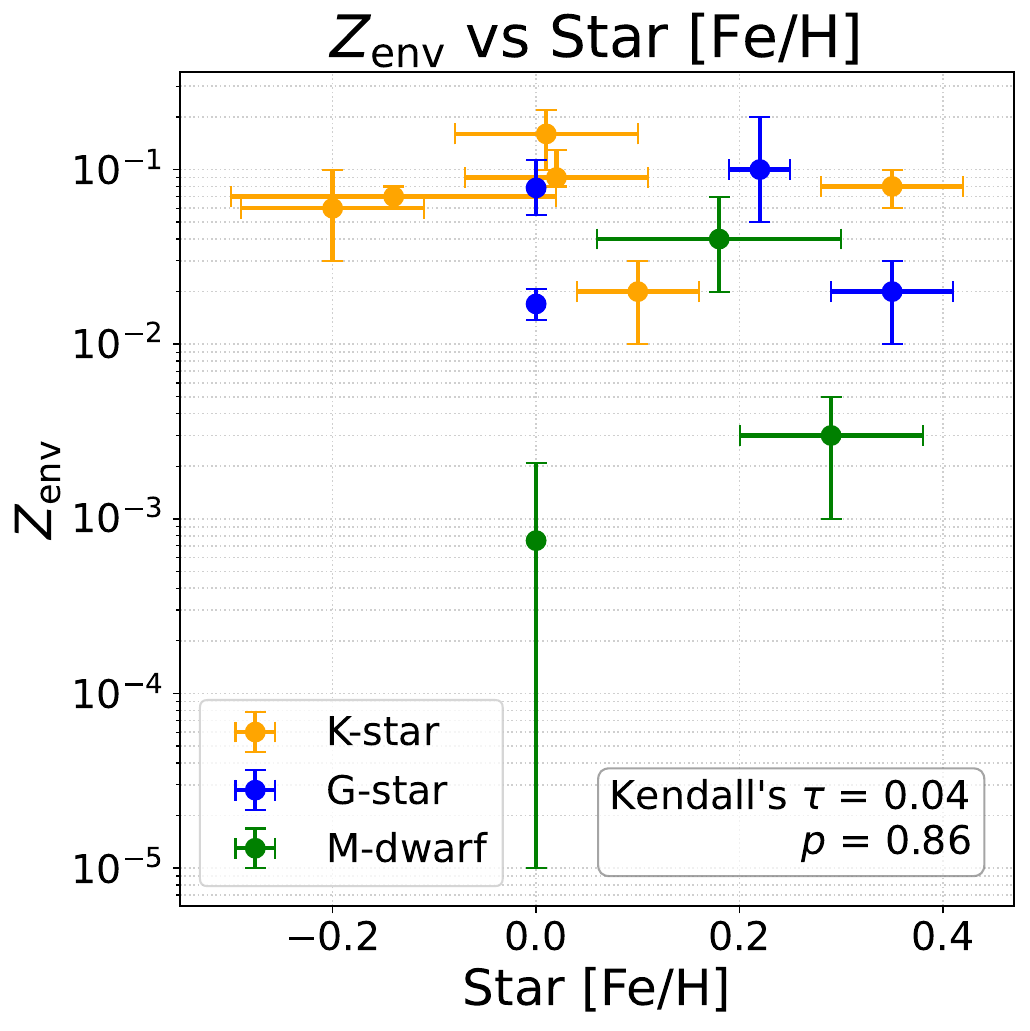}\includegraphics[width=0.33\textwidth]{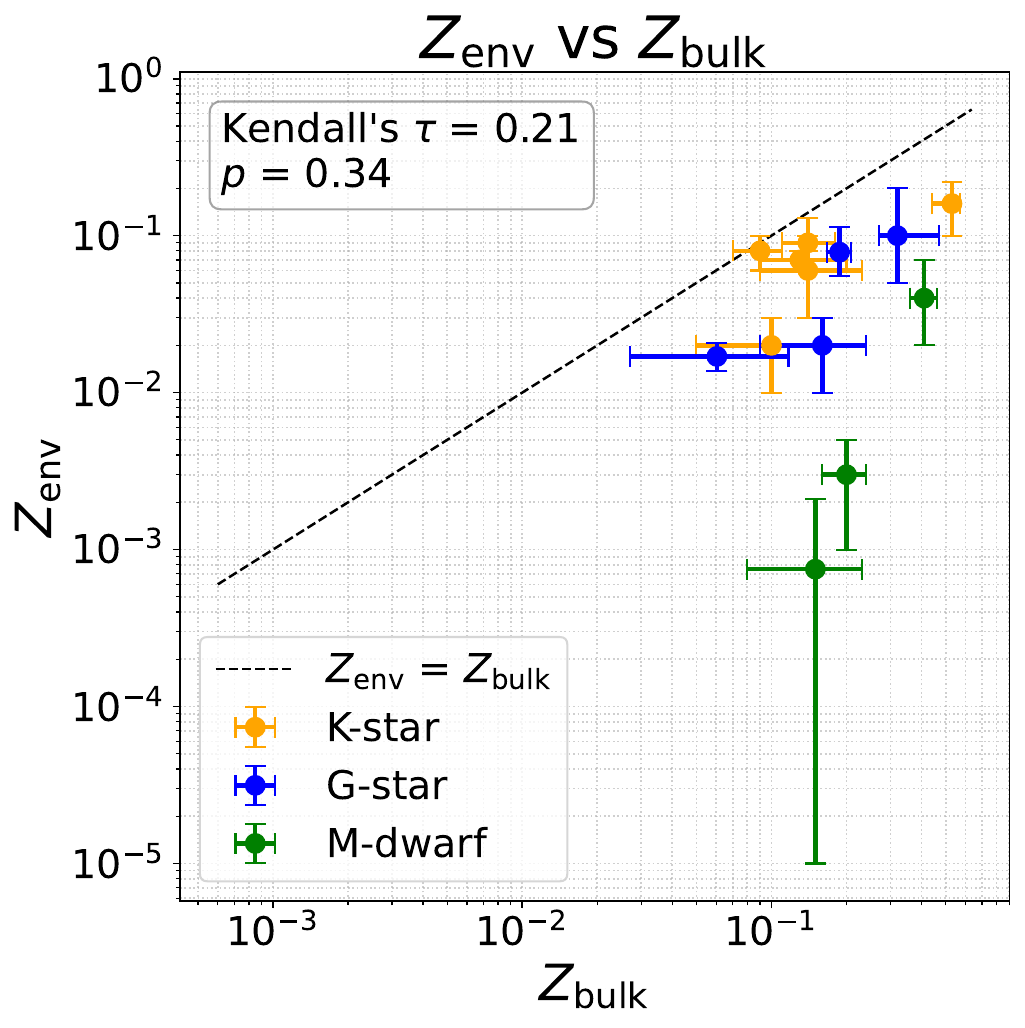}
  \caption{Exploratory analysis of the relation of a few key parameters in our planetary sample, including Jupiter and Saturn. Left: Anti-correlation identified between envelope metallicity and planetary mass ($p=0.0092$). Center: Envelope versus stellar metallicity. Right: Envelope versus bulk metallicity. Envelope metallicity errors are from the atmospheric measurements, while errors on the bulk metallicity are from the interior retrievals. Interestingly, there is no significant correlation between the envelope and stellar metallicity, nor the envelope and bulk metallicity; refer to the Discussion for further analysis.}
  \label{fig:three_side_by_side}
\end{figure*}

Although our sample is relatively small, we performed a set of simple statistical tests to investigate potential trends and correlations among key planetary parameters. The resulting relationships are shown as scatter plots in Figure \ref{fig:three_side_by_side}. We quantified the strength and significance of these correlations using Kendall's $\tau$ statistic and the associated $p$-values.

Notably, we find a statistically significant anti-correlation between the planetary mass and envelope metallicity. This trend is consistent with the previously suggested correlation between the bulk metallicity and the planet mass \citep[e.g.,][]{thorngren_massmetallicity_2016,muller_bulk_2025}. It is also expected within the core-accretion model of giant planet formation \citep{mizuno_formation_1980,pollack_formation_1996}, wherein a planet undergoing runaway gas accretion sweeps in hydrogen and helium far more efficiently than the concurrent delivery of solid materials, thereby diluting the envelope as the planet's mass increases. 

We did not find a statistically significant correlation between the bulk and envelope metallicity. Naively, one may expect a planet with a high metal budget should in principle have formed from a metal-rich material region in the protoplanetary disk which, in turn, would enrich the atmosphere. However, we caution against over-interpreting this result, since we once again stress that our planet sample is small, and more data is needed. Additionally, we also note that JWST measurements only probe the atmospheric composition at very low pressures and, therefore, potentially do not represent the deeper envelope.

We also did not find a correlation between the $Z_\text{env}$ and stellar metallicity [Fe/H].
%The standard expectation is that the star-forming reservoir of metals also sets the composition of the surrounding planet-forming disk, giving rise to some correlation between stellar and planetary metal abundances \citep{fischer_planet-metallicity_2005,guillot_correlation_2006}. 
The absence of a correlation between $Z_\text{env}$  and [Fe/H] does not contradict the established correlation between stellar metallicity and the bulk heavy-element content of giant planets \citep{guillot_correlation_2006,thorngren_massmetallicity_2016}. Instead, it reflects the distinction between atmospheric and bulk metallicity in giant planets, and is consistent with the key conclusion of this study, i.e., that atmospheric metallicity and bulk metallicity in giant planets can differ significantly. 
%studies showing that stellar metallicity is more closely related to the planet's total heavy-element inventory \citep{Guillot2006, MillerFortney2011, Thorngren2016}.
%One potential explanation for the absence of such a trend could be the fundamental physical difference in what constitutes "metallicity" when comparing stellar versus planetary contexts; a star's metallicity is dominated by iron-peak elements (hence "[Fe/H]"), while water and silicates are the dominant metals in giant planets.m
%As such, host-star iron abundance may not an ideal proxy for the volatile-rich material available to form the planetary envelope.

\section{Discussion}\label{sec:discussion}

The planet sample in this work is clearly limited by the scarcity of JWST-measured planets, and is made smaller still as we restrict it to planets within a certain mass range and equilibrium temperatures.
Nevertheless, 8 out of 11 planets we analyzed had a $Z_\text{env}/Z_\text{bulk} < 0.5$, with a median ratio $Z_\text{env} / Z_\text{bulk}=0.25^{+0.30}_{-0.22}$. Also, 10 out of 11 planets were inconsistent with a fully mixed planet (a mixing ratio of one) to within one sigma. This suggests that the planets in our sample have some degree of compositional layering and are not fully mixed.

As already described in Section \ref{sec:methods}, \textsc{GASTLI} assumes that the planets are made of three layers. Therefore, finer structural details of the interior --- namely, how this variation may materialize (i.e. via compositional gradients vs stratified layers) --- is beyond the scope of this work, as both options plausibly explain the observed inhomogeneous distribution of metals. However, since we generally find a mixing ratio significantly different from unity, our qualitative conclusion that these planets are not fully mixed is robust despite this simplification.

The planets in our sample orbiting M-dwarf stars stand out for their strongest compositional variation: their comparatively metal-poor atmospheres suggest protoplanetary disks around M-dwarf stars may have a smaller reservoir of metal-rich material available to form planets.
Disks formed around M-dwarf stars are observed to be systematically less massive than those of solar-type stars \citep{pascucci_steeper_2016,ansdell_alma_2017}---plausibly, the available heavy elements coalesced to form the interior, leaving very few metals to form the envelope.
The result is a metal-depleted envelope with a substantial heavy-element reservoir in the deeper interior, which would explain the $Z_\text{env}/Z_\text{bulk}$ observed for planets with M-dwarf hosts.

A noteworthy outlier in our sample is the nearly fully-mixed WASP-69 b, i.e. the planet exhibiting the most homogeneity between its envelope and bulk metal mass fraction ($Z_\text{env}/Z_\text{bulk} = 0.89$).
Interestingly, WASP-69 b is known to be undergoing atmospheric mass loss \citep{nortmann_ground-based_2018}; theoretically, as the outer, hydrogen-rich layers of the envelope are stripped away, the heavier species previously occupying the deeper interior may be dredged upward \citep{madhusudhan_exoplanetary_2016, hazra_atmospheric_2025}.
This unique pollution of the envelope would produce a higher degree of similarity between the bulk and envelope metallicity than is seen in non-evaporating planets, and could potentially explain its status as the best-mixed planet in our sample.

\section{Conclusions}\label{sec:conclusions}
We combined JWST-derived atmospheric metallicities with measured masses, radii, and system ages to infer the bulk metallicities of eleven warm giant exoplanets using \textsc{GASTLI}, and compared the envelope and bulk heavy-element mass fractions in order to assess how well mixed these planets are. This constitutes the first such comparison for a sample of exoplanets. Our main conclusions are:

\begin{itemize}
    \item Across the entire sample, we find $Z_{\rm env} < Z_{\rm bulk}$, with $Z_{\rm env}/Z_{\rm bulk}$ spanning approximately $0.02$--$0.90$. The majority of planets occupy the moderately stratified regime, $0.10 \lesssim Z_{\rm env}/Z_{\rm bulk} \lesssim 0.50$, indicating that their heavy elements are preferentially concentrated below the observable envelope. Notably, none of the planets in our sample exhibits an envelope enriched in heavy elements relative to its bulk composition ($Z_{\rm env}/Z_{\rm bulk} > 1$). This systematic offset between atmospheric and bulk compositions provides evidence that substantial compositional gradients may be a common feature of warm giant exoplanets.
    
    \item Our results indicate that incomplete mixing is prevalent among the warm giants currently accessible to JWST characterization. This has important implications for the interpretation of atmospheric abundances: atmospheric metallicity should not, in general, be assumed to trace the bulk heavy-element content of a giant planet. Treating a measured [M/H] as representative of the planet as a whole would therefore systematically underestimate its total heavy-element inventory, potentially leading to significant biases in inferences about planetary formation and evolution.
    
    \item Jupiter and Saturn occupy the same range of mixing ratios as the exoplanet sample. The similarity between the Solar System giants and warm giant exoplanets suggests that the compositional gradients inferred for Jupiter and Saturn may not be unique to our Solar System. Instead, dilute cores and partially mixed interiors may represent a common outcome of giant planet formation and evolution, with important implications for the processes governing the accretion, redistribution, and long-term transport of heavy elements \cite{Helled2022}.
    
    \item We identify a statistically significant anti-correlation between planetary mass and envelope metallicity. This trend is consistent with expectations from core-accretion models, in which rapid runaway gas accretion increasingly dilutes the heavy-element content of the growing envelope relative to the material accumulated during the earlier stages of formation. If confirmed by larger and more homogeneous samples, this relation could provide an observational constraint on the connection between giant planet mass, accretion history, and the efficiency with which heavy elements are incorporated into and redistributed within planetary interiors.
\end{itemize}

Taken together, our results provide evidence that the atmospheric compositions of warm giant exoplanets do not, in general, directly reflect their bulk heavy-element content. Across our sample, we find $Z_{\rm env} < Z_{\rm bulk}$, with most planets occupying a moderately stratified regime in which heavy elements are preferentially concentrated below the observable envelope. This has important implications for the interpretation of atmospheric metallicities and for our understanding of giant planet formation and evolution: atmospheric abundances alone provides only a partial view of a the planetary bulk composition, and substantial compositional gradients may be a common feature of giant planet interiors. The fact that Jupiter and Saturn occupy a similar range of envelope-to-bulk metallicity ratios further suggests that partially mixed or dilute interiors may not be peculiar to the Solar System, but could represent a common outcome of giant planet formation \cite{Valletta2020,Helled2023}. The statistically significant anti-correlation we identify between planetary mass and envelope metallicity additionally points to a connection between present-day atmospheric composition and accretion history, consistent with the increasing dilution of heavy elements during runaway gas accretion.

However, these conclusions should be interpreted in light of several important limitations. Our sample is currently small and heterogeneous, and the adopted atmospheric metallicities are derived from retrievals employing different codes, datasets, and assumptions; for individual planets, these estimates can differ by factors of several. Moreover, JWST transmission and emission spectra probe relatively shallow atmospheric pressures that may not be representative of the composition of the deep envelope, particularly when processes such as condensation and photochemistry modify the observable atmospheric abundances. On the modelling side, the true uncertainties in $Z_{\rm bulk}$ are likely larger than implied by our statistical errors alone, owing to systematic uncertainties associated with the adopted equation of state (EOS), atmospheric boundary conditions, and assumed interior structure. These limitations highlight the need for caution when interpreting the observed atmospheric-to-bulk metallicity ratios and the correlations identified here.

Nevertheless, the results presented here demonstrate the potential of combining atmospheric characterization with interior modelling to probe the otherwise inaccessible composition of giant planets. Expanding and homogenizing the sample is essential to assess whether the trends identified in this work are robust and representative of the broader population. Continued JWST characterization of warm giant exoplanets, together with the forthcoming Ariel mission, will extend these comparisons to a substantially larger and more homogeneous population, enabling the relationship between atmospheric composition, planetary mass, and bulk heavy-element enrichment to be tested with greater statistical power. In parallel, homogeneous reanalyses of atmospheric spectra and next-generation interior models that self-consistently account for composition gradients, realistic equations of state, atmospheric boundary conditions, and evolutionary effects will be crucial for reducing the dominant systematic uncertainties.

Ultimately, combining increasingly precise atmospheric observations with physically consistent models of planetary interiors offers a powerful route toward connecting what we can observe at the tops of planetary atmospheres with the formation histories and hidden interiors of giant planets. As the observational sample grows, the comparison between atmospheric and bulk compositions may provide a new empirical window into the efficiency of solid and gas accretion, the redistribution of heavy elements, and the development of compositional gradients during planetary evolution. The emerging synergy between JWST, Ariel, and advanced interior models therefore has the potential to transform atmospheric metallicity from a measurement of the observable atmosphere into a powerful diagnostic of how giant planets form, evolve, and acquire their compositional diversity.

%We note that our sample is small and heterogeneous. The adopted atmospheric metallicities are derived from retrievals using different codes and assumptions, and estimates for a given planet can differ by factors of several. Moreover, JWST transmission and emission spectra probe relatively shallow atmospheric pressures, which may not be representative of the composition of the deep envelope, particularly when processes such as condensation and photochemistry modify the observable atmospheric abundances. 
%On the modelling side, the true uncertainty in $Z_{\rm bulk}$ is likely larger than indicated by our statistical uncertainties, as it also includes systematic uncertainties associated with the adopted equation of state (EOS), atmospheric boundary conditions, and assumed interior structure. 

%Expanding and homogenizing the sample would substantially improve the robustness of our conclusions. Continued JWST characterization of warm giant exoplanets, together with the forthcoming Ariel mission, will extend this comparison to a larger and more homogeneous population, enabling the trends identified here to be tested more rigorously. On the theoretical side, interior models that self-consistently account for composition gradients, combined with a homogeneous reanalysis of the available atmospheric spectra, could further reduce the associated systematic uncertainties. 

\begin{acknowledgements}
    We acknowledge support from the Swiss National Science Foundation (SNSF) grant \texttt{\detokenize{200020_215634}} and the National Centre for Competence in Research ‘PlanetS’ supported by SNSF. This research used data from the NASA Exoplanet Archive, which is operated by the California Institute of Technology, under contract with the National Aeronautics and Space Administration under the Exoplanet Exploration Program. The interior and evolution models were calculated with \textsc{GASTLI} \citep{acuna_gastli_2024}. Extensive use was made of the \textsc{Python} packages \textsc{Jupyter} \citep{jupyter}, \textsc{Matplotlib} \citep{Hunter2007}, \textsc{NumPy} \citep{harris2020array}, \textsc{SciPy} \citep{virtanen_scipy_2020}, and \textsc{emcee} \citep{foreman-mackey_emcee_2013}.
\end{acknowledgements}

\bibliographystyle{aa}
\bibliography{references, references-2}

\begin{appendix}

\section{Stellar parameters}\label{sec:appendix_1}

\begin{table}[ht!]
\renewcommand{\arraystretch}{1.1}
\renewcommand{\tabcolsep}{3pt}
\caption{Stellar parameters}                 % title of Table
\label{table:3}    % is used to refer this table in the text
\centering                        % used for centering table
\begin{tabular}{lcccr}      % centered columns (4 columns)
\hline\hline               % inserts double horizontal lines
Star & Spec. Type & Mass ($M_{\odot}$) & $Z_{*}/Z_\odot$ & $T_{\text{eff}}$ (K) \\         % table heading
% \hline 
% \multicolumn{5}{c}{\it GASTLI } \\
\hline
% inserts single horizontal line
  HAT-P-26 & K1 & 0.36$^{+0.02}_{-0.01}$ & 0.01$^{+0.09}_{-0.09}$ & 5102 $\pm$20 \\
  WASP-107 & K6 & 0.68$^{+0.02}_{-0.02}$ & 0.02$^{+0.09}_{-0.09}$ & 4425$\pm$ 70 \\
  TOI-3984 A b & M4 & $0.49\pm0.02$ & $0.18^{+0.12}_{-0.12}$ & $3476\pm88$ \\
  TOI-199 & G9 & 0.94$^{+0.003}_{-0.01}$ & 0.22$^{+0.03}_{-0.03}$ & 5255$\pm$12\\
  HAT-P-18 & K2 & 0.77$^{+0.03}_{-0.03}$ & 0.10$^{+0.06}_{-0.06}$ & 4870$\pm$50 \\
  HAT-P-12 & K5 & 0.69$^{+0.03}_{-0.02}$ & -0.20$^{+0.09}_{-0.09}$ & 4665$\pm$45 \\
  WASP-69 & K5 & 0.83$^{+0.05}_{-0.05}$ & 0.35$^{+0.07}_{-0.07}$ & 4792$\pm$158 \\
  % HATS-6 & M1 & 0.60$^{+0.02}_{-0.02}$ & 0.20$^{+0.10}_{-0.10}$ & 3770$\pm$100 \\
  HATS-75 & M0 & 0.60$^{+0.01}_{-0.01}$ & 0.29$^{+0.09}_{-0.09}$ & 3812$\pm$79 \\
  WASP-80 & K7 & 0.57$^{+0.05}_{-0.05}$ & -0.14$^{+0.16}_{-0.16}$ & 4150$\pm$100 \\
  % TOI-3714 & M2 & $0.53^{+0.02}_{-0.02}$ & $0.1^{+0.1}_{-0.1}$ & $3660\pm90$ \\
  TOI-5205 & M4 & 0.39$^{+0.02}_{-0.02}$ & 0.00$^{+0.00}_{-0.00}$ & 3430$\pm$54 \\
  HD-80606 & G5 & 1.05$^{+0.02}_{-0.02}$ & 0.35$^{+0.06}_{-0.06}$ & 5565$\pm92$ \\
%  TOI-5293 A & M3 & 0.48$^{+0.01}_{-0.01}$ & -0.03$^{+0.12}_{-0.12}$ & 3586$\pm$ 88 \\
\hline
\vspace{0.1 mm}
\end{tabular}
\caption*{References: \textbf{HAT-P-26}: \cite{mancini_gaps_2022} (EXOFASTv2); \cite{2011ApJ...728..138H} - \textbf{WASP-107}: \cite{piaulet_wasp-107bs_2021} - \textbf{TOI-3984 A}: \cite{canas_toi-3984_2023} - \textbf{TOI-199}: \cite{hobson_toi-199_2023} - \textbf{HAT-P-18}: \cite{esposito_gaps_2014} -  \textbf{HAT-P-12}: \cite{mancini_gaps_2018} - \textbf{WASP-69}: \cite{allart_nirps_2025} - \textbf{HATS-75}: \cite{jordan_hats-74ab_2022} - \textbf{WASP-80}: \cite{triaud_wasp-80b_2013} - \textbf{TOI-5205}: \cite{kanodia_toi-5205_2023} - \textbf{HD-80606}: \cite{rosenthal_california_2021}}
\end{table}

% \subsection{Comparison to other interior models}\label{sec:appemndix_2}
% Here we compare our results to other attempts to compute the bulk and envelope metallicities of some of the exoplanets in the sample, summarized in Table \ref{littable}. We acknowledge the significant uncertainties relating to atmospheric models and contradicting retrievals. There exists much nuance between methods, and retrieval results depend strongly on the adopted modeling framework.
% For example, the decision to include (or exclude) clouds, hazes, or chemical equilibrium assumptions (e.g. radiative-convective thermal equilibrium vs. disequilibrium chemistry) can have a significant impact on the predicted atmospheric composition \citep[see][for an analysis of this impact on a few of the more common retrieval methods]{barstow_comparison_2020}.

% \cite{thorngren_connecting_2019} uses a combination of RV and TTV follow-up.
% i.e. no JWST data; \cite{acuna-aguirre_bulk_2025} uses panchromatic transmission + emission spectra from HST+JWST; JR6 considers equilibrium chemistry + no additional heat sources. 
% \cite{canas_gems_2025}.

% \input{tables/table3}

\end{appendix}

\end{document}